\documentclass[twocolumn,superscriptaddress]{revtex4-1}

\usepackage{amsmath}
\usepackage{amssymb}
\usepackage{amsthm}
\usepackage{graphicx}
\usepackage{mathbbol}
\usepackage{mathtools}
\usepackage{url}

\newcommand{\bra}[1]{\langle #1 |}
\newcommand{\ket}[1]{| #1 \rangle}

\newcommand{\ketbra}[2]{| #1 \rangle \langle #2 |}
\newcommand{\expect}[1]{\langle #1 \rangle}

\begin{document}

\title{Faithful measure of Quantum non-Gaussianity via quantum relative entropy}

\author{Jiyong Park}
\affiliation{School of Basic Sciences, Hanbat National University, Daejeon 34158, Korea}
\author{Jaehak Lee}
\affiliation{Department of Physics, Texas A \& M University at Qatar, P.O. Box 23874, Doha, Qatar}
\affiliation{School of Computational Sciences, Korea Institute for Advanced Study, Seoul 02455, Korea}
\author{Kyunghyun Baek}
\affiliation{Department of Physics, Texas A \& M University at Qatar, P.O. Box 23874, Doha, Qatar}
\affiliation{School of Computational Sciences, Korea Institute for Advanced Study, Seoul 02455, Korea}
\author{Se-Wan Ji}
\affiliation{National Security Research Institute, Daejeon 34044, Korea}
\author{Hyunchul Nha}
\email{hyunchul.nha@qatar.tamu.edu}
\affiliation{Department of Physics, Texas A \& M University at Qatar, P.O. Box 23874, Doha, Qatar}
\affiliation{School of Computational Sciences, Korea Institute for Advanced Study, Seoul 02455, Korea}
\begin{abstract}

We introduce a measure of quantum non-Gaussianity (QNG) for those quantum states not accessible by a mixture of Gaussian states in terms of quantum relative entropy. Specifically, we employ a convex-roof extension using all possible mixed-state decompositions beyond the usual pure-state decompositions. We prove that this approach brings a QNG measure fulfilling the properties desired as a proper monotone under Gaussian channels and conditional Gaussian operations. As an illustration, we explicitly calculate QNG for the noisy single-photon states and demonstrate that QNG coincides with non-Gaussianity of the state itself when the single-photon fraction is sufficiently large.
\end{abstract}

\maketitle

\section{Introduction}
Quantum mechanics provides a profound basis for many distinguished information processing protocols which cannot be achieved in the classical world, such as quantum computation \cite{bib:Chuang}, quantum teleportation \cite{bib:teleportation}, and quantum cryptography \cite{bib:cryptography}. Those quantum protocols have been developed also using continuous-variables (CVs) that can be usually described in terms of quasiprobability distributions like Glauber-Sudarshan P-function or the Wigner function in phase space \cite{Barnett, Book Zubairy}. A wide range of states like the coherent and the squeezed states are categorized as the so-called Gaussian states whose quasi-probability distributions take a Gaussian form and whoser statistical properties are completely characterized by their first-order moments (amplitudes) and the second-order moments (covariances). Gaussian states and Gaussian operations are widely employed in many CV protocols due to their experimental feasibility in laboratory with their compact mathematical description \cite{bib:Gaussian}. Nevertheless, there exist numerous no-go theorems within Gaussian regime, which prevent Gaussian operations from performing important tasks such as universal quantum computation \cite{computation1,computation2}, quantum error correction \cite{errorcorrection}, and entanglement distillation \cite{distillation1,distillation2,distillation3}, also addressed recently in the framework of Gaussian resource theories \cite{Lami}. In such tasks, non-Gaussian states and non-Gaussian operations become essential resources.

In this respect, it is of crucial importance to identify quantum non-Gaussian states that cannot be produced by Gaussian resources and their statistical mixtures. Furthermore, it may provide a valuable framework and a novel insight into related studies to characterize quantum non-Gaussianity (QNG) under a proper quantitative measure. 
In a closely related context, several studies have investigated to quantify non-Gaussianity (NG) of quantum states \cite{NGmeasure1, NGmeasure2, NGmeasure3}, which only represents the departure of a given state from Gausian states. In particular, it was shown that relative entropy of NG exhibits important properties, for example, monotonicity under Gaussian channels \cite{NGmeasure4}. However, the measure is not convex because the set of Gaussian states is not convex. There indeed exist non-Gaussian states which can be simply generated using Gaussian operations and classical randomness, for example, a mixture of two different coherent states $ ( \ket{\alpha}\bra{\alpha} +  \ket{-\alpha}\bra{-\alpha} ) /2 $. These states, a simple mixture of Gaussian states, can be generated without quantum non-Gaussian operations and they are thus not suitable to perform quantum information tasks which requires genuinely quantum non-Gaussian resources.

Recently some works have devoted to ruling out Gaussian mixtures and detecting genuinely quantum non-Gaussian states, i.e. $\rho\ne\sum_ip_i\rho_{G,i}$, where each component state $\rho_{G,i}$ is a Gaussian state. Though a number of criteria have been developed to assess quantum non-Gaussian states \cite{QNGdetect1, QNGdetect2, QNGdetect3,QNGdetect4,QNGdetect5,QNGdetect6,QNGdetect7,QNGdetect8,QNGdetect9,QNGdetect10}, a faithful measure of quantum non-Gaussianity has not been reported yet. Recent studies in Refs. \cite{Takagi, Albarelli} have remarkably adopted the Wigner negativity as a measure of QNG, which is a monotone under Gaussian protocols including classical mixing. However it is actually not a faithful measure because it cannot detect quantum non-Gaussian states with positive Wigner function, e.g. a highly noisy single-photon state $p|0\rangle\langle0|+(1-p)|1\rangle\langle1|$ with $p>0.5$. A recent work by Takagi {\it et al.} suggests that every resource state can generally provide an operational advantage in view of subchannel discrimination even including Quantum non-Gaussian states with positive Wigner functions \cite{TTakagi}. Therefore, it seems necessary to come up with a QNG measure that can broadly and faithfully assess quantum non-Gaussian states.

In this work, we propose a convex-roof measure of QNG based on quantum relative entropy. Our QNG measure is faithful because it always gives a positive value whenever a state cannot be described as a Gaussian mixture. We prove that our measure satisfies properties as a proper measure of QNG including convexity, additivity, and monotonicity under Gaussian channels and conditional Gaussian operations. Furthermore, we illustrate how to explicitly evaluate QNG for a noisy single-photon state. We find that its QNG coincides with its NG if the single-photon fraction is large enough.

\section{QNG measure via relative entropy and its properties}
\subsection{Non-Gaussianity}
We first start with the notion of non-Gaussianity (NG). For a given mixed state $\rho$, one may define its NG  in terms of quantum relative entropy with reference to its Gaussified state $\rho_G$ having the same first-order moments (average) and second-order moments (covariance) \cite{NGmeasure2}. 
That is, 
${\cal N}[{\rho}]\equiv S(\rho||\rho_G)$ where $S(\rho||\sigma)\equiv-{\rm Tr} \{\rho\log \sigma\}+{\rm Tr} \{\rho\log \rho\}$ is quantum relative entropy. In particular, due to  $-{\rm Tr} \{\rho\log \rho_G\}=-{\rm Tr} \{\rho_G\log \rho_G\}$, we have the relation $S(\rho||\rho_G)=S(\rho_G)-S(\rho)$, which highlights the fact that a Gaussian state among all states with the same covariance matrix possesses a maximal entropy leading to the nonnegativity of the defined NG \cite{bib:Marian}.

\subsection{Quantum non-Gaussianity}
We are here interested in quantum non-Gaussianity (QNG) of states, which cannot be represented by a mixture of Gaussian states,  namely, $\rho\ne\sum_ip_i\rho_G^i$. There can be several approaches to quantify the degree of QNG and we use the convex-roof extension of NG defined above. That is, for a given state $\rho$, its QNG can be measured as 
\begin{eqnarray}
{Q}[{\rho}]\equiv{\rm min}_{\{p_i,\rho_i\}}\sum_ip_iS(\rho_i|| \rho_{i,G})
\end{eqnarray}
where the minimization is taken over all possible decompositions of $\rho=\sum_ip_i\rho_i$. Note that this generalization includes the usual decomposition into pure-states only, $\rho=\sum_ip_i|\Psi_i\rangle\langle\Psi_i|$, e.g. in \cite{Albarelli}. By further allowing decompositions into mixed states, we may obtain a lower degree of QNG for a given state. We will illustrate it later by pointing out a range of noisy single-photon states whose QNG is given by a genuinly mixed-state decomposition.

We prove the following properties of the above-defined QNG.

{\bf N0}: {\it  QNG is nonnegative.}

---This is obvious by its definition, as the relative entropies, and thus their average, are nonnegative.

{\bf N1}: (faithfulness) {\it QNG is strictly positive if and only if the state is not a mixture of Gaussian states.}

---This can also be readily seen. If $\rho=\sum_ip_i\rho_{G,i}$, its QNG is then zero due to the decomposition with Gaussian component states only. On the other hand, if the QNG is zero, it also means that the given state is a mixture of Gaussian states since any single non-Gaussian component state, if any, would give a strictly positive NG, leading to a positive QNG.

{\bf N2}: (convexity) {\it QNG is convex with respect to state mixing, i.e.} ${Q}[{\lambda\rho_1+(1-\lambda)\rho_2}]\le \lambda {Q}[{\rho_1}]+ (1-\lambda) Q[{\rho_2}]$.

---Proof: Let $\rho_1=\sum_ip_i\rho_i$ and $\rho_2=\sum_jq_j\sigma_j$ be the decompositions for their respective QNGs. Since $\sum_i\lambda p_i\rho_i+\sum_j(1-\lambda)q_j\sigma_j$ is one possible decomposition of the state $\lambda\rho_1+(1-\lambda)\rho_2$, we have by definition 
\begin{eqnarray}
&&Q[{\lambda\rho_1+(1-\lambda)\rho_2}]\nonumber\\
&&\le \sum_i\lambda p_iS(\rho_i||\rho_{i,G})+\sum_j(1-\lambda)q_jS(\sigma_j||\sigma_{j,G})\nonumber\\&&=\lambda Q[{\rho_1}]+ (1-\lambda) Q[{\rho_2}].
\end{eqnarray}

{\bf N3}: {\it QNG is invariant under Gaussian unitary operations.}

---Proof: For any fixed decomposition $\rho=\sum_ip_i\rho_i$, a Gaussian unitary operation leads to $\rho'=U_G\rho U_G^\dag=\sum_ip_iU_G\rho_i U_G^\dag$. We also note that the relative entropy of each component NG is invariant under unitary operation, $S(\rho_i||\rho_{i,G})=S(U\rho_iU^\dag||U\rho_{i,G}U^\dag)$ and that the Gaussification of state commutes with Gaussian unitary operations. The latter property means that $U_G\rho_{i,G} U_G^\dag$ is the Gaussified state of $\rho'=U_G\rho_{i} U_G^\dag$. Therefore, $\sum_ip_iS(\rho_i|| \rho_{i,G})$ is invariant under Gaussian unitary operations and so is QNG.

{\bf N4}: {\it QNG is not increasing under Gaussian channels.}

---Proof: 
\begin{eqnarray}
Q[\rho]&&={\rm min} \sum_ip_iS(\rho_i|| \rho_{i,G})\nonumber\\&&\ge \sum_ip_iS({\cal E}_G(\rho_i)|| {\cal E}_G(\rho_{i,G}))\ge Q[{{\cal E}_G(\rho)}],
\end{eqnarray}
where the first inequality is due to the contraction property of relative entropy under an arbitrary quantum channel. Note again that ${\cal E}_G(\rho_{i,G})$ is equivalent to the Gaussified state of ${\cal E}_G(\rho_i)$ and that $ \sum_ip_i{\cal E}_G(\rho_i)$ is one of possible decompositions of ${\cal E}_G(\rho)$, which leads to the second inequality
in Eq. (2).

{\bf N5}: {\it QNG is not increasing on average under conditional Gaussian maps.}

For its proof, we first introduce two preliminary tools.
 
{\it Preliminary 1}---Takagi and Zhuang in \cite{Takagi} have identified a general conditional Gaussian map as the one attaching an ancillary (multi-mode) vacuum to the system followed by a global unitary Gaussian operation and homodyne detection. The conditional map results from implementing a Gaussian map conditioned on the measurement outcome. That is, 
with $\rho_{SE}=U_G|0\rangle\langle0|\otimes\rho_sU_G^\dag$, we obtain $\rho'=\sum_k|k\rangle\langle k|\otimes\rho_k=\sum_kp_k|k\rangle\langle k|\otimes{\tilde \rho}_k$, where $\rho_k=\langle k|\rho_{SE}|k\rangle$ is an unnormalized state conditioned on the homodyne outcome $k$ with $p_k={\rm Tr} \rho_k$.
The final conditional map reads $\rho''=\sum_kp_k|k\rangle\langle k|\otimes{\cal E}_G^k({\tilde \rho}_k)$.\\

{\it Preliminary 2}---For two mixed states $\rho=\sum_j p_j^{(1)}|j\rangle\langle j|\otimes\rho_j$ and $\sigma=\sum_j p_j^{(2)}|j\rangle\langle j|\otimes\sigma_j$ where $|j\rangle$'s are orthonormal states for subsystem A,  the relative entropy $S(\rho||\sigma)$ turns out to be 
\begin{eqnarray}
S(\rho||\sigma)=H(p^{(1)}||p^{(2)})+\sum_j p_j^{(1)}S(\rho_j||\sigma_j), 
\end{eqnarray}
where $H$ is the Shannon relative entropy. Using these properties, we have the following proof.

---Proof:
We first note that the QNG of $\rho_{SE}=U_G|0\rangle\langle0|\otimes\rho_sU_G^\dag$ is the same as that of $\rho_s$, since neither addding an ancillary Gaussian state nor a unitary Gaussian operation changes QNG. 
 Let $\rho_{SE}=\sum_ip_i\rho_i$ be the decomposition yielding its QNG, i.e. $Q[{\rho_s}]=Q[{\rho_{SE}}]=\sum_ip_iS(\rho_i|| \rho_{i,G})$, where $\rho_i$ belongs to a larger Hilbert space of \{SE\}. 

We may introduce a further extended state of  $\rho_{SE}$ as $\rho_{SEE'}=\sum_ip_i|i\rangle\langle i|\otimes\rho_i$ where $\rho_{SE}={\rm Tr}_{E'}\{\rho_{SEE'}\}$ and $|i\rangle$'s are orthonormal basis states for $E'$. 
With its ``Gaussified" version $\sigma_{SEE'}=\sum_ip_i|i\rangle\langle i|\otimes\rho_{i,G}$, we have $Q[{\rho_{SE}}]=S(\rho_{SEE'}||\sigma_{SEE'})$  due to Preliminary 2, that is, expressed in terms of the relative entropy of the total states without decompositions.

Let us now take a homodyne measurement with basis $|k\rangle$ on subsystem E for the two states $\rho_{SEE'}$ and $\sigma_{SEE'}$. We then obtain 
\begin{eqnarray}
\rho_{SEE'}\rightarrow \rho_{SEE'}'=&& \sum_{i,k}p_i|i\rangle\langle i|\otimes |k\rangle\langle k| \otimes \langle k|\rho_{i}|k\rangle\nonumber\\
=&& \sum_{i,k}p_ip_{k|i}|i\rangle\langle i|\otimes |k\rangle\langle k| \otimes {\tilde\rho}_{k|i},
\end{eqnarray}
where ${\tilde\rho}_{k|i}$ is the normalized state obtained on the measurement outcome $k$ starting with the state $\rho_i$ and $p_{k|i}={\rm Tr} \langle k|\rho_{i}|k\rangle$ is the corresponding conditional probability. The product $p_ip_{k|i}\equiv p_{ik}$ defines a joint probability as such. Similarly, we obtain the state after measurement for $\sigma_{SEE'}$, however, the conditional probability $p_{k|i}^G={\rm Tr} \langle k|\rho_{i,G}|k\rangle$ is not necessarily the same as $p_{k|i}={\rm Tr} \langle k|\rho_{i}|k\rangle$. Nevertheless, with 
\begin{eqnarray}
\sigma_{SEE'}\rightarrow \sigma_{SEE'}'= && \sum_{i,k}p_i|i\rangle\langle i|\otimes |k\rangle\langle k| \otimes \langle k|\rho_{i,G}|k\rangle\nonumber\\
=&& \sum_{i,k}p_ip_{k|i}^G|i\rangle\langle i|\otimes |k\rangle\langle k| \otimes {\tilde\rho}_{k|i,G},
\end{eqnarray}
and since a measurement on a partial system is a CP map (its action is actually to eliminate all off-diagonal elements in the subsytem), we have $S(\rho_{SEE'}||\sigma_{SEE'})\ge S(\rho_{SEE'}'||\sigma_{SEE'}')$.
Using the Preliminary 2 again, the latter quantity is given by $H(p_{ik}||p_{ik}^G)+\sum_{i,k}p_{ik}S({\tilde\rho}_{k|i}||{\tilde\rho}_{k|i,G})\ge\sum_kp_kS_k$, where $H\ge0$ is used.
We have here defined the marginal probability $p_k=\sum_ip_{ik}$ and $S_k=\frac{1}{p_k}\sum_i p_{ik} S({\tilde\rho}_{k|i}||{\tilde\rho}_{k|i,G})$. 

Noting that $\rho_k=\frac{1}{p_k}\sum_ip_{ik}{\tilde\rho}_{k|i}$ is the state of system conditioned on the measurement outcome $k$ on $E$, we have therefore proved  
$Q[\rho]=Q[{\rho_{SE}}]=S(\rho_{SEE'}||\sigma_{SEE'})\ge S(\rho_{SEE'}'||\sigma_{SEE'}')\ge\sum_kp_kS_k\ge\sum_k p_k Q[{\rho_k}]\ge \sum_k p_k Q[{{\cal E}_G^k(\rho_k)}]$.


\section{Case of noisy single-photon states}
In the previous section, we have demonstrated that our entropic QNG measure fulfills desirable properties as a proper measure of quantum non-Gaussianity.
Operationally, we may interpret our measure as quantifying the minimum required non-Gaussian resources to prepare a given quantum non-Gaussian state. 
We have specifically introduced the convex-roof extension adopting mixe-state decompositions beyond the usual pure-state decompositions to define the degree of QNG. 
One may then be interested to know if there exist quantum non-Gaussian states whose QNG is given strictly by a mixed-state decomposition not by a pure-state decomposition. We illustrate it by an example of noisy single-photon states with the explicit calculation of their QNG based on our approach. Before that, we remark on the case of pure non-Gaussian states.

\subsection{pure states}
If the state is pure, $\rho=|\Psi\rangle\langle\Psi|$, the state itself is the only possible decomposition of it. Therefore, its QNG coincides with its NG, $ Q[{\rho}]={\cal N} [{\rho}]$.

\subsection{Noisy single-photon state}
We now consider the case of mixed states. 
Specifically, we obtain the QNG of a noisy single-photon state, i.e. $p \ketbra{1}{1} + (1-p) \ketbra{0}{0}$, as follows. 

(i) To begin with, we obtain the non-Gaussianity, not QNG yet, of a noisy single-photon state in a general form of $\rho = p \ketbra{1}{1} + (1-p) \ketbra{0}{0} + re^{i\theta} \ketbra{0}{1} + re^{-i\theta} \ketbra{1}{0}$. For this state, we have $\expect{\hat{a}} = \expect{\hat{a}^{\dag}}^{*} = r e^{-i \theta}$, $\expect{\hat{a}^{2}} = \expect{(\hat{a}^{\dag})^{2}} = 0$ and $\expect{\hat{a} \hat{a}^{\dag}} = \expect{\hat{a}^{\dag} \hat{a}} + 1 = p + 1$, which yield $\expect{\hat{q}} = \sqrt{2} r \cos \theta$, $\expect{\hat{p}} = - \sqrt{2} r \sin \theta$, $\expect{\hat{q}^{2}} = \expect{\hat{p}^{2}} = \frac{1}{2} + p$ and $\expect{\hat{q}\hat{p}+\hat{p}\hat{q}}=0$ where $\hat{q} = \frac{\hat{a}+\hat{a}^{\dag}}{\sqrt{2}}$ and $\hat{p} = \frac{\hat{a}-\hat{a}^{\dag}}{ \sqrt{2}i}$ are two orthogonal quadrature amplitudes. The covariance matrix of $\rho$ is then given by
	\begin{equation}
		\Gamma = \begin{pmatrix} \frac{1}{2} + p - 2r^{2}\cos^{2}\theta & 2r^{2}\sin\theta\cos\theta \\ 2r^{2}\sin\theta\cos\theta & \frac{1}{2} + p - 2r^{2}\sin^{2}\theta \end{pmatrix},
	\end{equation}
where the covariance matrix elements are defined as $\Gamma_{ij} = \frac{1}{2} \expect{\hat{x}_{i} \hat{x}_{j} + \hat{x}_{j} \hat{x}_{i}} - \expect{\hat{x}_{i}} \expect{\hat{x}_{j}}$ with $\hat{x}_{1} = \hat{q}$ and $\hat{x}_{2} = \hat{p}$. It determines the quantum entropy of the reference Gaussian state $\rho_{G}$ as
	\begin{equation}
		S ( \rho_{G} ) = ( \bar{n}_{\mathrm{th}} + 1 ) \log ( \bar{n}_{\mathrm{th}} + 1 ) - \bar{n}_{\mathrm{th}} \log \bar{n}_{\mathrm{th}},
	\end{equation}
where
	\begin{equation}
		\bar{n}_{\mathrm{th}} = \sqrt{\det \Gamma} - \frac{1}{2} = \sqrt{(\frac{1}{2}+p)(\frac{1}{2}+p-2r^{2})} - \frac{1}{2}.
	\end{equation}
The non-Gaussianity $\rho$ is thus given by
	\begin{align}\label{eq:NGNSPT}
		{\cal N}[{\rho}] = &S ( \rho_{G} ) -S ( \rho ) \nonumber\\=& ( \bar{n}_{\mathrm{th}} + 1 ) \log ( \bar{n}_{\mathrm{th}} + 1 ) -  \bar{n}_{\mathrm{th}} \log \bar{n}_{\mathrm{th}} \nonumber \\
		& + \lambda_{+} \log \lambda_{+} + \lambda_{-} \log \lambda_{-},
	\end{align}
where $\lambda_{\pm} = \frac{1}{2} \pm \sqrt{(\frac{1}{2}-p)^{2}+r^{2}}$ are the eigenvalues of $\rho$.
Note that the NG of the state $\rho$ is independent of the phase $\theta$, which is indeed due to the invariance property under Gaussian unitary operations, particularly phase rotation in this case, i.e. ${\cal N}[\rho]={\cal N}[e^{i{\hat n}\theta}\rho e^{-i{\hat n}\theta}]$. 

\begin{figure}[!ht]
		\includegraphics[scale=0.45]{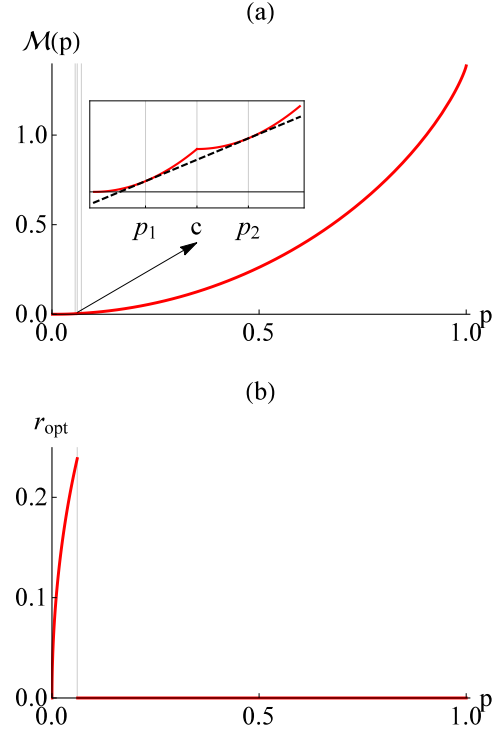}
          	\caption{(a) minimum $\mathcal{M} (p)$ defined in Eq. (11) as a function of the single photon fraction $p$ and (b) corresponding optimimal parameter $r_{\mathrm{opt}}$ for minimum $\mathcal{M} (p)$.}
		\label{fig:OP}
	\end{figure}

(ii) From the non-Gaussianity in Eq.~\eqref{eq:NGNSPT}, we may find the minimum of $NG_{\rho}$ among all states for a fixed $p$ as
	\begin{equation}
		\mathcal{M} (p) \equiv \min_{r} {\cal N}[{\rho}], 
	\end{equation}
which can be obtained by solving 
	\begin{align}
		\frac{d}{dr} {\cal N}[{\rho}] = & 4 r \bigg\{ \frac{\tanh^{-1} (2\lambda_{+}-1)}{2\lambda_{+}-1} - \frac{1+2p}{2 \bar{n}_{\mathrm{th}}} \tanh^{-1} \frac{1}{2 \bar{n}_{\mathrm{th}}} \bigg\} \nonumber \\
		=  & 0,
	\end{align}
and comparing the extremal values. We plot the minimum $\mathcal{M} (p)$ and the corresponding optimal parameter $r_{\mathrm{opt}}$ as a function of $p$ in Fig.~\ref{fig:OP}. The minimum NG is given by a partially mixed state ($0 < r_{\mathrm{opt}} < \sqrt{p(1-p)}$) and a maximally mixed state ($r_{\mathrm{opt}} = 0$) for $p \lesssim 0.062$ and $p \gtrsim 0.062$, respectively.

(iii) Using the above result, we obtain the QNG of $\rho_p = p \ketbra{1}{1} + (1-p) \ketbra{0}{0}$ as follows. Given a state $\rho_p$, our task is to find a decomposition yielding $Q[\rho_p]={\rm min}_{\{f_k,\rho_k\}}\sum_kf_k{\cal N}[\rho_k]$ among all decompositions $\rho_p=\sum_kf_k\rho_k$. In particular, we let $\rho_k$ be the state with single-photon fraction $p_k$ thus satisfying the constraint $p=\sum_kf_kp_k$. The idea of optimization here is to find values $p_k$ with the constraint $p=\sum_kf_kp_k$ to have a minimum $\sum_kf_k{\cal M}[p_k]$, where ${\cal M} [p]$ is the function whose values are shown in Fig. 1. 

This optimization actually corresponds to the lower convex envelope of $\mathcal{M} (p)$ defined by
\begin{eqnarray}
\breve{\mathcal{M}}(p) \equiv \sup \{ f(p) | \mbox{$f$ is convex and $f \leq \mathcal{M}
$ in [0,1]} \}, \nonumber\\
\end{eqnarray}
 which is obtained as follows.
Investigating $\mathcal{M}^{\prime \prime} (p)$, we find that $\mathcal{M} (p)$ itself is convex on the two intervals $[0,c]$ and $[c,1]$ individually with $c \simeq 0.062$, but not in the whole interval (red solid curve in the inset of Fig.1 (a)). Then, we may construct the lower convex envelope by finding a line tangent to $\mathcal{M} (p)$ in both intervals [black dashed line in the inset of Fig. 1 (a)]. If there exists a solution to the equation
	\begin{equation}
		\mathcal{M}^{\prime} (p_1) ( p_2 - p_1 ) + \mathcal{M} (p_1) = \mathcal{M} ( p_2),
	\end{equation}
together with the condition $\mathcal{M}'(p_1) = \mathcal{M}'(p_2)$, the line is tangent to $\mathcal{M} (p)$ in both intervals. Indeed we find the solution $p_1 \simeq 0.0559$ and $p_2 \simeq 0.0701$, respectively.
 Therefore, we obtain the QNG of $\rho_p = p \ketbra{1}{1} + (1-p) \ketbra{0}{0}$ as 

\begin{align} \label{eq:QNGNSPT}
		 Q[{\rho_p}] 
		 =
		\begin{cases}
			\displaystyle \mathcal{M} (p) & \mbox{for $0 \leq p \leq p_1$,} \\
			\displaystyle \frac{p-p_1}{p_2-p_1} \mathcal{M} (p_2) + \frac{p_2-p}{p_2-p_1} \mathcal{M} (p_1) & \mbox{for $p_1 \leq p \leq p_2$,} \\
			\displaystyle \mathcal{M} (p) & \mbox{for $p_2 \leq p \leq1$,}
		\end{cases}
	\end{align}
where $p_1 \simeq 0.0559$ and $p_2 \simeq 0.0701$. 

From the above analysis, we can also identify an optimal decomposition of $\rho_p$ readily. For $p \geq p_2$ we have $\breve{\mathcal{M}} (p) = {\cal N}[{\rho}]$, which means that the state $\rho_p$ itself is the optimal decomposition attaining minimum convex roof QNG. This is a clear example for which the mixed-state decomposition becomes optimal rather than the pure-state decomposition. For $p \leq p_1$, the equal mixture of two optimal states $\rho_{\pm}^{p} = p \ketbra{1}{1} + (1-p) \ketbra{0}{0} \pm r_{\mathrm{opt}} ( \ketbra{0}{1} + \ketbra{1}{0} )$ achieves the bound. For the remaing case, i.e., $p_1 \leq p \leq p_2$,
the optimal decomposition becomes $\{ \rho_{+}^{p_1}, \rho_{-}^{p_1}, p_2 \ketbra{1}{1} + (1-p_2) \ketbra{0}{0} \}$ with the probability distribution $\{ \frac{1}{2} \frac{p_2-p}{p_2-p_1}, \frac{1}{2} \frac{p_2-p}{p_2-p_1}, \frac{p-p_1}{p_2-p_1} \}$.

\section{Discussion}
We have proposed a faithful measure of quantum non-Gaussianity adpoting quantum relative entropy. Specifically, we have introduced a convex-roof extension of non-Gaussianity using all possible mixed-state decompositions beyond the typical pure-state decompositions. This enables us to come up with properties desired as a proper measure of QNG including convexity and monotonicity under Gaussian channels and conditional Gaussian operations. Our measure is faithful in that it strictly gives a positive value for an arbitrary quantum non-Gaussian state that cannot be represented as a mixture of Gaussian states.

As an illustration, we have studied the case of a noisy-single photon state, which is a practically important QNG resource for many applications like linear-optical quantum computation \cite{Milburn}. We have shown the procedures to identify its QNG rigorously, which may be extended to quantum non-Gaussian states with higher photon numbers. By doing so, we have clearly illustrated that there exist a range of quantum states for which QNG is given by a mixed-state decomoposition, not a pure-state one. Moreover, it turns out that the QNG actually coincides with NG if the single-photon fraction is sufficiently large.

Our measure of QNG may be interpreted as quantifying the minimum required non-Gaussian resource to produce a given quantum non-Gaussian state. Namely, it addresses a way of preparing different non-Gaussian states with a proper probability distribution such that the average of non-Gaussianity of each state becomes minimal to constitute the quantum non-Gaussian state under investigation. While this measure has its own merit, a more comprehensive study is still needed concerning the characterization of QNG in a full variety of physical contexts . There have been some investigations demonstrating the usefulness of non-Gaussian states and operations, e.g. the improvement of quantum entanglement \cite{ED1,ED2,ED3,ED4,ED5,ED6,ED7} and enhancement of performance in quantum teleportation and dense coding \cite{QT1,QT2,QT3,QT4}. However, there were only a few studies to comprehensively and critically identify the role of QNG in CV quantum information processing beyond the level of case studies \cite{NPJ}. For instance, it is an interesting question whether an arbitrary quantum non-Gaussian state, even though it possesses a positive-definite Wigner function, can be a critically useful resource to provide an advantage for practical quantum tasks. If so, what sort of QNG measure would appropriately address such criticality in a rigorous way? These and other related issues shall be investigated elsewhere.

\section*{acknowledgement}
This work is supported by an NPRP grant 8-751-1-157 from Qatar National Research Fund.
J.P. acknowledges support by the National Research Foundation of Korea (NRF) grant funded by the Korea government (MSIT) (NRF-2019R1G1A1002337). 
S.-W.J. acknowledges support by the R \&D Convergence Program of NST (National Research Council of Science and Technology) of Republic of Korea
(Grant No. CAP-18-08-KRISS).

\end{document}